\documentstyle[pre,aps]{revtex}

\font\tenmsb=msbm10 scaled\magstep 1
\font\sevenmsb=msbm7 scaled \magstep 1
\font\faivemsb=msbm5 scaled
\textfont\msbfam=\tenmsb
\scriptfont\msbfam=\sevenmsb
\scriptscriptfont\msbfam=\faivemsb\def\Bbb#1{{\fam\msbfam #1}}

\font\tengothic=eufm10 scaled\magstep 1
\font\sevengothic=eufm7 scaled\magstep 1\newfam\gothicfam
\textfont\gothicfam=\tengothic
\scriptfont\gothicfam=\sevengothic

\newcommand{\cX}{{\cal X}}
\newcommand{\cF}{{\cal F}}
\newcommand{\cH}{{\cal H}}
\newcommand{\cL}{{\cal L}}
\newcommand{\cD}{{\cal D}}
\newcommand{\cP}{{\cal P}}
\newcommand{\cA}{{\cal A}}
\newcommand{\cR}{{\cal R}}
\newcommand{\vp}{\varphi}
\newcommand{\be}{\begin{equation}}
\newcommand{\ee}{\end{equation}}
\newcommand{\ep}{\varepsilon}
\newcommand{\ra}{\rightarrow}
\newcommand{\al}{\alpha}
\newcommand{\bt}{\beta}
\newcommand{\om}{\omega}
\newcommand{\dgr}{\dagger}
\newcommand{\bS}{{\bf S}}
\newcommand{\ba}{{\bf a}}

\begin{document}

\draft

\title{Quantifying Entanglement Production of Quantum Operations}
\author{V.I. Yukalov}
\address{Bogolubov Laboratory of Theoretical Physics, \\
Joint Institute for Nuclear Research, Dubna 141980, Russia}

\maketitle

\begin{abstract}

The problem of entanglement produced by an arbitrary operator is formulated
and a related measure of entanglement production is introduced. This measure
of entanglement production satisfies all properties natural for such a
characteristic. A particular case is the entanglement produced by a density
operator or a density matrix. The suggested measure is valid for operations
over pure states as well as over mixed states, for equilibrium as well as
nonequilibrium processes. Systems of arbitrary nature can be treated,
described either by field operators, spin operators, or any other kind of
operators, which is realized by constructing generalized density matrices.
The interplay between entanglement production and phase transitions in
statistical systems is analysed by the examples of Bose-Einstein condensation,
superconducting transition, and magnetic transitions. The relation between
the measure of entanglement production and order indices is analysed.

\end{abstract}

\vskip 1cm


\section{Introduction}

Entanglement is the term used by Schr\"odinger with respect to the
superposition principle applied to composite systems [1--3]. If two
quantum particles have interacted, their state cannot be presented
as a tensor product of single-particle states, but it is entangled,
being a superposition of such products. The notion of entanglement
is now at the heart of such interrelated intriguing problems as quantum
measurement, quantum information processing, and quantum computing,
which have been expounded in several books and reviews [4--12].

In the literature, one distinguishes between the entanglement of quantum
states and the entanglement produced by quantum operators over unentangled
wave functions. Both these types of entanglement, to be well defined, require
the knowledge of quantifying characteristics. In the present paper, as follows
from its title, we shall consider only the second type of entanglement
produced by quantum operations. In quantifying the latter, one usually envokes
the characteristics measuring the entanglement of states. Because of this, it
is useful to briefly mention the problem of quantifying the entanglement of
quantum states, which, at the same time, would make clearer the difference
between these two types of entanglement.

Quantifying entanglement of quantum states, one usually deals with bipartite
systems. Several ways of measuring entanglement in such systems have been
suggested, the methods being based on the notions of either reduced or
relative entropy. Thus, {\it mutual information} is a linear combination of
the von Neumann reduced entropies [13,14]. The reduced entropies themselves
define the {\it entropy of entanglement}, which serves as a measure for {\it
entanglement of formation} [15,16]. Another measure of entanglement is
defined by minimizing over disentangled states the {\it Kullback-Leibler
distance}, yielding the {\it relative entropy of entanglement} [17--19].
A measure, not envolving the notion of entropy, could be introduced as the
number of maximally entangled pairs that can be purified from a given state,
which results in {\it entanglement of distillation} [15]. However, this
measure depends on the particular process of purification, and it is not
clear how to compute it in an efficient and unique way [17--19].

Entanglement of formation was also employed in the attempt of characterizing
the entanglement for mixed states [20], but it was concluded that this way
did not uniquely define mixed-state entanglement [20,21]. A suggestion for
measuring {\it covariance entanglement} in bipartite systems by means of
correlation functions squared was advanced in [22]. However, such a measure,
as is accepted by the authors themselves, does not possess all necessary
properties [8--12] to be considered as really a measure. One often connects
the existence of entanglement with the violation of the Bell inequalities,
which can be formulated for bipartite systems with both orthogonal and
nonorthogonal states [23]. This point of view is based on the Gisin theorem
[24], according to which any pure entangled state of two particles violates
a Bell inequality for two-particle correlation functions. Nevertheless,
there exist pure entangled $N>2$ qubit states that do not violate any
Bell inequality for $N$-particle correlation functions [25]. It seems
that correlation functions do not provide the best tool for characterizing
entanglement. At present, there is no such a general definition of
entanglement measure that would be valid for bipartite as well as
multipartite systems, for pure as well as mixed states, for equilibrium
as well as nonequlilibrium processes.

Another problem is to quantify the entanglement produced by quantum
operations on a given set of disentangled functions. The so produced
entanglement is termed as entangling power, entanglement capacity,
entanglement of evolution, entanglement generation, or entanglement
production [26--30]. In what follows, we shall employ the term {\it
entanglement production} as the most closely related to the meaning
of this notion, being the entanglement produced by an operator. To
quantify this type of entanglement, one usually resorts to the
combination of measures defined for the entanglement of states. Since
the latter are well defined only for pure bipartive systems, the
entanglement production is also usually considered for such systems.

The aim of the present paper is to introduce a general {\it measure of
entanglement production}, which could quantify the amount of entanglement
produced by an arbitrary operator on a given disentangled set. The
suggested approach can be applied to operators of any nature and to
any physical systems, whether pure or mixed, bipartite or multipartite,
equilibrium or not. Being justified for arbitrary operators, the
approach can straightforwardly be applied to a particular kind of
operators, as statistical or density operators.

\section{Disentangled and entangled functions}

First of all, it is necessary to give correct mathematical definitions
for the notions of entangled or, conversely, disentangled functions and
to specify the notations to be employed in what follows. From the very
beginning, a multipartite composite system is kept in mind, consisting
of an arbitrary number of subsystems enumerated by the index $i=1,2,\ldots,p$,
where $p=1,2,\ldots$ can be any integer. Subsystems are treated as
indivisible parts, because of which they could equivalently be called
particles. These can be distinguishable or indistinguishable, which does
not change the general mathematical structure, provided an appropriate
labelling of partite states is employed, e.g., by means of collective
mode labels or occupation numbers [31--33].

The {\it space of single-partite states} for each $i$-part is presented
by the Hilbert space
\be
\label{1}
\cH_i \equiv \overline\cL\{ |n_i>\} \; ,
\ee
being a closed linear envelope of a basis $\{|n_i>\}$ composed of
orthonormalized single-partite states $|n_i>$. Hence, any vector
$\vp_i\in\cH_i$ can be expanded over the basis $\{|n_i>\}$ as
\be
\label{2}
\vp_i = \sum_{n_i} \; a_{n_i}\; |n_i> \; .
\ee
The nature of the labels $n_i$ here is of no importance. The vector norm
$$
||\vp_i||_{\cH_i} \equiv \sqrt{(\vp_i,\vp_i)}
$$
in $\cH_i$ is defined through the associated scalar product $(\vp_i,\vp_i)$.

It is worth emphasizing that $|n>$ does not compulsorily mean a quantum state
of a given physical particle. This case is not excluded if particles are
distinguishable. But, if one deals with a system of indistinguishable
particles, $|n>$ should be understood as a single-particle mode.

The {\it space of composite-system states} is given by the $p$-fold tensor
product
\be
\label{3}
\cH \equiv \otimes_{i=1}^p \cH_i \; ,
\ee
which is identified [34] with the closed linear envelope
\be
\label{4}
\cH \equiv \overline{\cL}\{ | n_1\ldots n_p>\}
\ee
over a normalized $p$-particle basis $\{|n_1\ldots n_p>\}$. The latter may
be written as the tensor product
\be
\label{5}
|n_1\ldots n_p> \; = \otimes_{i=1}^p |n_i>
\ee
of the single-particle basis states. Any function $\vp\in\cH$ can be
presented as the sum
\be
\label{6}
\vp = \sum_{\{ n_i\} } \; c_{n_1\ldots n_p}\; |n_1\ldots n_p >
\ee
over the multiparticle basis $\{|n_1\ldots n_p>\}$, where $\{ n_i\}\equiv
n_1,n_2,\ldots, n_p$. The vector norm $||\vp||_\cH\equiv\sqrt{(\vp,\vp)}$
in $\cH$ is generated by the related scalar product $(\vp,\vp)$.

Note that it is not compulsory to deal with the total tensor-product space (3).
In some cases, because of physical restrictions, additional selection rules
may be superimposed on the admissible states of $\cH$. This is, e.g., the
case of systems composed of identical particles, whose quantum states are
to be either symmetrized or antisymmetrized, according to whether the
particles are bosons or fermions. Then the space of admissible states is
reduced to a subspace of space (3). In what follows, we shall keep in mind
the possibility of such additional restriction rules. For short, we shall
continue denoting by $\cH$ a subspace of the space structure (3), implying
that all necessary selection rules are taken into account. In the presence
of the latter, the structure (3) is termed the incomplete tensor product
[35--39].

The multiparticle space (3) contains the whole variety of different states,
from which one should separate out {\it disentangled states} or factor
states that are presentable as the tensor products $\otimes_{i=1}^p\vp_i$.
A collection $\{ f\}$ of all possible factor states $f$ forms the {\it
disentangled set}
\be
\label{7}
\cD \equiv \left\{ f=\otimes_{i=1}^p\vp_i \; , \;\;
\vp_i\in\cH_i\right\} \; ,
\ee
which is a subset of the Hilbert space (3). A function $f\in\cD\subset\cH$
can be written as
\be
\label{8}
f=\otimes_{i=1}^p \; \sum_{n_i} \; a_{n_i}\; |n_i> \; ,
\ee
where expansion (2) is taken into account. For any two vectors $f$ and
$f'$ from $\cD$, the scalar product is
\be
\label{9}
(f,f') =\prod_{i=1}^p (\vp_i,\vp_i') \; .
\ee
This generates in $\cD$ the vector norm
\be
\label{10}
||f||_\cD = \prod_{i=1}^p ||\vp_i||_{\cH_i} \; .
\ee
The compliment $\cH\setminus\cD$ to $\cD$ forms the set of {\it entangled states}.

To illustrate in an explicit way the principal difference of disentangled states
from entangled ones, let us consider a bipartite system, with $p=2$, and let the
single-particle states be two-dimensional. Then, writing, for compactness, $1$
and $2$ instead of $n_1$ and $n_2$, for state (6), we have
$$
\vp =c_{11}|11>\; + \; c_{12}|12>\; + \; c_{21}|21>\; + \; c_{22}|22> \; ,
$$
while for the factor state (8), we get
$$
f=a_1b_1|11> \; + \; a_1 b_2|12> \; + \; a_2b_1|21>\; + \; a_2b_2|22>\; .
$$
As is evident, the state $\vp\in\cH$ is more general than $f\in\cD$. The space
$\cH$ contains the entangled states, such as $c_{12}|12>+c_{21}|21>$ or
$c_{11}|11>+c_{22}|22>$ that in no way can be reduced to the factor states
$f\in\cD$. In general, no entangled state can be presented as a product of
single-particle states.

\section{Measure of Entanglement Production}

The multiparticle space (3) contains both entangled and factor states. For the
time being, we follow the abstract terminology of Sec. II, implying no physical
applications that will be treated later. An abstract mathematical level of
consideration provides the best way for making transparent what actually is
entanglement production and how to measure it.

The term entanglement production as such means an action that transforms
disentangled states into entangled ones. A transformation can be described by
the action of an operator. Thus, one may investigate entanglement produced by
different operators. Let $A$ be an arbitrary linear bounded operator acting on
the tensor-product space $\cH$. A complete theory of linear bounded operators,
defined on the tensor-product spaces, can be found in Refs. [35--39].

The norm of a linear operator $A$ on $\cH$ can be given by
\be
\label{11}
||A||_\cH =\sup_{||\vp||_\cH=1}||A\vp||_\cH =
\sup_{||\vp||_\cH=||\vp'||_\cH=1} |(\vp,A\vp')| \; .
\ee
For a bounded operator, the norm is finite. If the operator $A$ is Hermitian,
then
$$
||A||_\cH = \sup_{||\vp||_\cH=1} |(\vp,A\vp)| \qquad (A^+=A) \; .
$$

Let us introduce the projector $\cP_\cD$, which projects the total space (3)
onto its subset given by the disentangled set (7), so that
\be
\label{12}
\cP_\cD\cH =\cD \; ,
\ee
with the standard properties of projecting operators
$$
\cP_\cD^2 =\cP_\cD \; , \qquad \cP_\cD^+=\cP_\cD \; , \qquad
||\cP_\cD||_\cH = 1 \; .
$$
The projector in Eq. (12) is nonlinear, therefore the equality for its norm
has to be understood as a definition. And let us define the norm of $A$ on
$\cD$ as
\be
\label{13}
||A||_\cD \equiv ||\cP_\cD A\cP_\cD||_\cH \; .
\ee
This, in view of the structure of the disentangled set (7), can be presented
as
\be
\label{14}
||A||_\cD = \sup_{||f||_\cD=||f'||_\cD=1}\; |(f,Af')| \; .
\ee

An operator $A$, acting on $f\in\cD$, generally, transforms $f$ to an
$\vp\in\cH$. That is, an operator $A$, in general, entangles the factor
states. The operator $A$, having the property $A\cD\subset\cH\setminus\cD$ will
be termed {\it entangling operator}.

Similarly to the existence of entangled and factor states, there exist
entangling operators and nonentangling ones. The latter should, clearly, have
the structure of a direct product $\otimes_{i=1}^p A_1^i$ of single-particle
operators $A_1^i$ acting on $\cH_i$. Let the algebra of all linear bounded
operators on $\cH$ be denoted by $\cA\equiv\{ A\}$. And let us separate out
from this algebra a subset $\cA^\otimes\equiv\{\cA^\otimes\}\subset\cA$ of
nonentangling, or product, operators having the structure of a product
$\otimes_{i=1}^p A_1^i$. Thus, by construction,
\be
\label{15}
\cA\cD\subset\cH \; , \qquad \cA^\otimes\cD\subset\cD \; .
\ee
Analogously to the projection of $\cH$ onto $\cD$ by means of the projector
$\cP_\cD$ given in Eq. (12), we may denote the reduction of $\cA$ to
$\cA^\otimes$ with the help of a projector $\cP_\otimes$, such that
\be
\label{16}
\cA^\otimes =\cP_\otimes(\cA) \; ,
\ee
which is a superoperator acting on the Hilbert space of linear operators,
with the properties
$$
\cP_\otimes^2 =\cP_\otimes \; , \qquad \cP_\otimes^+ =\cP_\otimes \; ,
\qquad ||\cP_\otimes||_\cD = 1
$$
being valid. The equality for the norm has to be understood as a definition.
An explicit construction of a product operator $A^\otimes\subset
\cA^\otimes$, associated with a given operator $A\subset\cA$, can be done as
follows. Let us define a reduced single-particle operator $A_1^i$ on $\cH_i$
as
\be
\label{17}
A_1^i \equiv C_i\; {\rm Tr}_{\{\cH_{j\neq i}\} } \; A \; ,
\ee
where the trace runs over all $\cH_j$ except the case $j=i$. The set of
constants $C_i$ is chosen so that to satisfy the normalization condition
\be
\label{18}
{\rm Tr}_\cH \; A = {\rm Tr}_\cD \; A^\otimes \; .
\ee
In this way, we obtain the {\it product operator}
\be
\label{19}
A^\otimes \equiv \;
\frac{{\rm Tr}_\cH A}{{\rm Tr}_\cD\otimes_{i=1}^p A_1^i} \;
\otimes_{i=1}^p\; A_1^i \; ,
\ee
where
$$
{\rm Tr}_\cD\; \otimes_{i=1}^p\; A_1^i = \prod_{i=1}^p
{\rm Tr}_{\cH_i}\; A_1^i \; .
$$
Thus, from a given multiparticle space (3), it is possible to separate out
the disentangled set (7) and, similarly, for an arbitrary operator $A$, one
can put into correspondence the product operator (19). That is, there exist
entangled and disentangled states as well as entangling and nonentangling
operators.

Let us be interested in the entanglement produced by an operator $A$ on $\cH$.
How could we measure the resulting entanglement production? It would seem
natural that we should somehow compare the actions of the given operator $A$
and its nonentangling counterpart $A^\otimes$. But what quantity should be
defined for this purpose? Following the common ideology, one should construct
a sort of entropy, comparing, say, ${\rm Tr}_\cH A\ln A$ and
${\rm Tr}_\cD A^\otimes\ln A^\otimes$. However, this is not the best way.
The pivotal idea, we shall follow, is the observation that the {\it norm} of
an operator characterizes a kind of order associated with this operator [40].
For instance, invoking the norm and trace of reduced density operators, or
reduced density matrices, it is possible to define the {\it density order
indices} [41]. Generalizing the latter, one can define the {\it operator
order indices} [40] for arbitrary operators. These order indices provide
a complete classification for different types of order, long-range and
mid-range, off-diagonal and diagonal, because of which they are applicable
for describing both phase transitions and crossover phenomena.

Entanglement, in some sense, is also a characteristic of order (or disorder).
Hence it should be well characterized by an operator norm. To this end, we
introduce the {\it measure of entanglement production}
\be
\label{20}
\ep(A) \equiv \log\; \frac{||A||_\cD}{||A^\otimes||_\cD}
\ee
generated by the given operator $A$. Here the logarithm can be taken
with respect to any base that would be convenient, e.g., to the base 2.

As follows from definition (20), the entanglement-production measure quantifies
the {\it amount of entanglement produced by an operator} $A$ over a set $\cD$,
because of which this measure should, in general, be denoted as $\ep(A,\cD)$.
When working with a fixed set $\cD$, one may, for simplicity, shorten the
notation writing $\ep(A) = \ep(A,\cD)$. Quantity (20) satisfies all natural
properties that are compulsory for being really a measure:

\vskip 2mm

{\bf 1}. {\it Measure is semipositive}.

This is evident by construction, since
$$
||A^\otimes||_\cD = ||\cP_\otimes(A)||_\cD
\leq ||A||_\cD \; ,
$$
because of which
\be
\label{21}
\ep(A) \geq 0 \; .
\ee

\vskip 2mm

{\bf 2}. {\it Measure is continuous}.

This implies the following. Assume that for any operator $A$ of $\cH$ there
exists a family $\{ A(t)\}$ of operators $A(t)$ parameterized with $t\in\Bbb{R}$,
so that $A(t)\ra A$ as $t\ra 0$ in the sense of the norm convergence
$$
||A(t)||_\cD \ra ||A||_\cD \qquad (t\ra 0) \; .
$$
If so, then for the measure (20) one has
\be
\label{22}
\ep(A(t))\ra \ep(A) \qquad (t\ra 0) \; .
\ee
The nature of the parameter $t$ can be arbitrary. In physical applications,
this may be time, temperature, density, interaction parameters, and so on.

\vskip 2mm

{\bf 3}. {\it Measure is zero for a nonentangling operator}.

A nonentangling operator has the form of $A^\otimes$. As is obvious, for
$A=A^\otimes$,
\be
\label{23}
\ep(A^\otimes)= 0 \; .
\ee
In particular, there is no self-entanglement of a single-particle system,
when $A=A_1=A^\otimes$.

\vskip 2mm

{\bf 4}. {\it Measure is additive}.

Let now $\{ A_\nu\}$ be a set of copies of $A$, with $\nu=1,2,\ldots$, so
that $A=\otimes_\nu A_\nu$. Such a case may happen, e.g., in treating
heterophase systems [42]. Then $A^\otimes=\otimes_\nu A_\nu^\otimes$ and one
gets
$$
||A||_\cD = \prod_\nu \; ||A_\nu||_\cD \; , \qquad
||A^\otimes||_\cD = \prod_\nu \; ||A_\nu^\otimes||_\cD \; .
$$
From here, it follows that
\be
\label{24}
\ep\left ( \otimes_\nu\; A_\nu\right ) = \sum_\nu\; \ep(A_\nu) \; .
\ee

\vskip 2mm

{\bf 5}. {\it Measure is invariant under local unitary operations}.

Such operations are described by a set $\{ U_i\}$ of unitary operators $U_i$
on $\cH_i$, with $U_i^+U_i=1$ and $i=1,2,\ldots,p$. Using the properties of
the operator norm and trace, it is easy to show that
\be
\label{25}
\ep\left (\otimes_{i=1}^p \; U_i^+\; A\otimes_{i=1}^p\; U_i\right ) =
\ep(A) \; .
\ee

Thus, the entanglement-production measure (20) can be introduced for any
operator. The requirement that the latter be bounded can be relaxed in the
following way. Assume that $A$ defined on $\cH$ is unbounded. Introduce a
restricted space $\cR_N$ such that the norm $||A||_{\cR_N}$ is finite and
$\cH$ can be treated as the inductive limit $\cR_N\ra\cH$ as $N\ra\infty$.
Here $N$ is not necessarily the number of particles, but may be any labelling
number. Following the procedure described above, one can separate out from
the space $\cR_N$ the disentangled set $\cD_N$, whose inductive limit is
$\cD=\lim_{N\ra\infty}\;\cD_N$. Then the entanglement measure, instead
of Eq. (20), can be defined as
$$
\ep(A) \equiv \lim_{N\ra\infty} \; \log\;
\frac{||A||_{\cD_N}}{||A^\otimes||_{\cD_N}} \; .
$$
In particular, this can correspond to the thermodynamic limit.

\section{Pure State Entanglement}

Entanglement production, as is explained above, can be generated by any
operator. In physical applications, one often implies that this is due
to the action of a von Neumann operator. There are two types of density
operators, pure-state and mixed-state. A density operator $\hat\rho_N$
for an ensemble of $N$ particles is termed a pure-state operator when
it is an idempotent operator, such that $\hat\rho_N^2=\hat\rho_N$.
Below, we shall consider several examples of calculating the
entanglement-production measure (20) generated by pure-state density
operators.

\subsection{Einstein-Podolsky-Rosen states}

These states provide a classical example of entangled states. They
correspond to a bipartite system composed of two-dimensional parts, so
that such a state can be presented as
\be
\label{26}
|EPR>\; \equiv \frac{1}{\sqrt{2}}\; (|12>\; \pm \; |21> ) \; .
\ee
The related density operator is
\be
\label{27}
\hat\rho_{EPR} \; \equiv |EPR><EPR| \qquad \left ({\rm Tr}_\cH
\hat\rho_{EPR} = 1\right ) \; ,
\ee
with $\cH=\cH_1\otimes\cH_2$. Reduced single-particle density operators,
according to definition (17), have the form
$$
\hat\rho_1^i =\; \frac{1}{2}\; (|1><1| + |2><2| ) \; ,
$$
being given on $\cH_i$, with $i=1,2$. The product operator (19) is
$\hat\rho^\otimes_{EPR}=\hat\rho_1^1\otimes\hat\rho_1^2$. Calculating
the norm (13), we have for the density operator (27)
$$
||\hat\rho_{EPR}||_\cD  = ||\hat\rho_1^i||_{\cH_i} = \frac{1}{2} \; ,
$$
while for the product operator,
$$
||\hat\rho_{EPR}^\otimes||_\cD = ||\hat\rho_1^i||_{\cH_i}^2 =
\frac{1}{4} \; .
$$
Thence, the entanglement-production measure (20) is
\be
\label{28}
\ep(\hat\rho_{EPR}) =\log 2 \; .
\ee
Assuming the logarithm to the base $2$, one has $\ep(\hat\rho_{EPR})=1$.

\subsection{Bell states}

Similarly to the previous case, it is easy to find the
entanglement-production measure for the density operator formed by the
Bell states
\be
\label{29}
|B>\; \equiv \frac{1}{\sqrt{2}}\; (|11>\; \pm \; |22>) \; ,
\ee
so that the corresponding density operator is
\be
\label{30}
\hat\rho_B = |B><B| \qquad
\left ( {\rm Tr}_\cH\; \hat\rho_B=1\right ) \; .
\ee
For the entanglement-production measure (20), this gives
\be
\label{31}
\ep(\hat\rho_B) =\log 2 \; .
\ee
Keeping in mind the logarithm to the base $2$ yields $\ep(\hat\rho_B)=1$.

Note that the Bell states can be treated as a two-particle generalization
of the single-particle Schr\"odinger cat state.

\subsection{Greenberger-Horne-Zeilinger states}

Such states, having the form
\be
\label{32}
|GHZ>\; \equiv \frac{1}{\sqrt{2}} \; (|11\ldots 1> \; \pm \;
|22\ldots 2> ) \; ,
\ee
can be considered as an $N$-particle generalization of the single-particle
Schr\"odinger cat state [43,44]. The associated density operator is
\be
\label{33}
\hat\rho_{GHZ} \equiv |GHZ><GHZ| \qquad
\left ({\rm Tr}_\cH\hat\rho_{GHZ}=1 \right ) \; .
\ee
Following the procedure of Sec. III, we find
$$
||\hat\rho_{GHZ}||_\cD = ||\hat\rho_1^i||_{\cH_i} = \frac{1}{2} \; ,
$$
while for the corresponding product operator,
$$
||\hat\rho_{GHZ}^\otimes||_\cD = \frac{1}{2^N} \; .
$$
From here,
\be
\label{34}
\ep(\hat\rho_{GHZ}) = (N-1)\log 2 \; .
\ee
Hence, for the logarithm to the base $2$, one gets $\ep(\hat\rho_{GHZ})=N-1$.
As it should be, there is no entanglement production for a single particle,
when $N=1$, and the case of two particles, when $N=2$, reduces to measures
(28) or (31).

\subsection{Multicat states}

The multiparticle Schr\"odinger cat states, or, for short, multicat states,
are sometimes also called the generalized $GHZ$ states. They have the form
\be
\label{35}
|MC> \; \equiv c_1|11\ldots 1> \; + \; c_2|22\ldots 2>\; ,
\ee
where $N$ two-dimensional parts are assumed and the coefficients are arbitrary
complex numbers satisfying the normalization $|c_1|^2+|c_2|^2=1$. The related
density operator is
\be
\label{36}
\hat\rho_{MC} \equiv |MC><MC| \qquad
\left ({\rm Tr}_\cH\hat\rho_{MC}=1 \right ) \; .
\ee
For the norm (13), we find
$$
||\hat\rho_{MC}||_\cD = ||\hat\rho_1^i||_{\cH_i} =\sup\{ |c_1|^2, \;
|c_2|^2\} \; .
$$
The entanglement-production measure (20) becomes
\be
\label{37}
\ep(\hat\rho_{MC}) = (1-N)\log \sup\{|c_1|^2,\; |c_2|^2\} \; .
\ee
Its value lies in the interval
\be
\label{38}
0\leq \ep(\hat\rho_{MC} ) \leq (N-1)\log 2 \; .
\ee
The maximal entanglement production occurs when $|c_1|^2=|c_2|^2=1/2$, which
goes back to the $GHZ$ states. And the entanglement production disappears
if $|c_i|=1$ for any $i=1,2$.

Multicat states can be realized for systems of particles with two internal
single-particle states, such as trapped ions subject to the action of resonant
laser beams [45,46] or Bose-condensed neutral atoms with an effective
interaction due to coherent Raman scattering [47]. Instead of internal
single-particle states, one can create collective nonlinear states by invoking
the resonant excitation of topological coherent modes in trapped Bose-Einstein
condensates [48--50]. Such two-level or two-mode states are usually accompanied
by atomic squeezing [50,51].

\subsection{Multimode states}

A natural generalization of the multicat states are the multimode states
describing a system of $N$ parts, each of which can be in one of $m$
different modes. Such a state reads
\be
\label{39}
|MM>\; \equiv \sum_n \; c_n |nn\ldots n> \; ,
\ee
where
$$
\sum_n \; |c_n|^2 = 1 \; , \qquad \sum_n \; 1  = m \; .
$$
The corresponding density operator is
\be
\label{40}
\hat\rho_{MM} \equiv |MM><MM| \qquad
\left ({\rm Tr}_\cH\hat\rho_{MM} = 1 \right ) \; .
\ee
For this case, the entanglement-production measure writes
\be
\label{41}
\ep(\hat\rho_{MM}) = ( 1 - N)\log\sup_n |c_n|^2 \; .
\ee
It varies in the range
\be
\label{42}
0\leq \ep(\hat\rho_{MM}) \leq (N-1)\log m \; .
\ee
The maximal entanglement production is reached when $|c_n|=1/m$, while
entanglement is absent if any of $|c_n|=1$. For the two-mode case, when
$m=2$, one returns to the multicat states. In general, states (39) are
related to coherent states with a fixed number of particles [40,52,53].

\subsection{Hartree-Fock states}

These states, typical of indistinguishable particles, have the structure
\be
\label{43}
|HF> \; \equiv \frac{1}{\sqrt{N!}}\; \sum_{sym} \; |12\ldots N> \; ,
\ee
where a symmetrized or antisymmetrized sum is assumed according to either
bosons or fermions are considered. Note that here it is implied that $N$
particles are in $N$ different single-particle states. The density operator
is
\be
\label{44}
\hat\rho_{HF} \; \equiv |HF><HF| \qquad
\left ({\rm Tr}_\cH\; \hat\rho_{HF} = 1 \right ) \; .
\ee
Calculating
$$
||\hat\rho_{HF}||_\cD = \frac{1}{N!} \; , \qquad
||\hat\rho_1^i||_{\cH_i} = \frac{1}{N} \; ,
$$
we find the entanglement-production measure
\be
\label{45}
\ep(\hat\rho_{HF}) =\log\; \frac{N^N}{N!} \; .
\ee
For $N=2$, this reduces to
$$
\ep(\hat\rho_{HF}) =\log 2 \qquad (N=2) \; ,
$$
as it should be as far as state (43) reduces to the $EPR$ state (26). And
for large $N$, one gets
$$
\ep(\hat\rho_{HF}) \simeq N\log e \qquad (N \ra \infty) \; .
$$

It is worth noting that in all examples considered above the limit
$$
\lim_{N\ra\infty} \; \frac{1}{N}\; \ep(\hat\rho_N) < \infty
$$
exists. The existence of such a limit is sometimes required as
a prerequisite property of any entanglement measure. In our case, the
existence of this limit is guaranteed by the property of additivity (24).

\section{Mixed state Entanglement}

Mixed states of physical systems are characterized by density operators
that are not idempotent, so that $\hat\rho^2_N\neq\hat\rho_N$. Since the
entanglement-production measure (20) is defined for arbitrary operators,
there is no principal problem of applying this definition to any density
operators, including the mixed-state density operators. When a statistical
operator is entangling, the related measure of produced entanglement must
be nonzero.

To concretize the aforesaid, let us consider the operator
$$
\hat\rho \equiv \frac{1}{2} \left ( |11><11| + |22><22|\right ) \; .
$$
This form corresponds to what one terms a separable statistical
operator. However, despite its simple form, the operator $\hat\rho$
produces entanglement, i.e., it is {\it entangling}. This is easy to
demonstrate by taking, for instance, a product function
$$
f=\sqrt{2}\; (\; |1>+|2> \;) \otimes (\; |1>+|2>\; )
$$
from the disentangled set $\cD$. The action of $\hat\rho$ on this
function gives
$$
\hat\rho\; f = \frac{1}{\sqrt{2}}\; ( |11>+|22>\; ) \; ,
$$
which is a Bell state, that is, a maximally entangled state. Here
$\hat\rho$ really does produce entanglement. Therefore, the related
measure of produced entanglement (20) has to be nonzero. The latter
can be easily verified following the general scheme. Thus, we have
$$
\hat\rho_i ={\rm Tr}_{\cH_{j\neq i}}\; \hat\rho =
\frac{1}{2}\; (\; |1><1| + |2><2|\; ) \; ,
$$
so that $\hat\rho^\otimes=\hat\rho_1\otimes\hat\rho_2$. Calculating
the norms
$$
||\hat\rho||_\cD  =\frac{1}{2}\; , \qquad ||\hat\rho_i||_{\cH_i}=
\frac{1}{2}\; , \qquad ||\hat\rho^\otimes||_\cD = \frac{1}{4} \; ,
$$
we find $\ep(\hat\rho)=\log 2$, as it should be for an entangling
operator $\hat\rho$.

\subsection{Quantum mechanics}

Let us consider mixed-state density operators obtained by taking partial
traces of pure-state quantum density operators. For instance, let us
consider the Hartree-Fock pure state density operator (44). Its partial
traces produce the mixed-state reduced density operators
\be
\label{46}
\hat\rho_{HF}^p \equiv {\rm Tr}_{\cH_{p+1}}
\ldots {\rm Tr}_{\cH_N} \rho_{HF} \; ,
\ee
with $p=1,2,\ldots N-1$. It is easy to find the norm
$$
||\hat\rho_{HF}^p||_\cD = \frac{(N-p)!}{N!} \; ,
$$
from where the entanglement-production measure (20) becomes
\be
\label{47}
\ep(\hat\rho^p_{HF}) = \log\;
\frac{(N-p)! N^p}{N!} \; .
\ee
For a large number of particles and finite $p$ this gives
\be
\label{48}
\ep(\hat\rho^p_{HF}) \simeq \; \frac{p(p-1)}{2N}\;
\log e \qquad (N\ra\infty) \; .
\ee
Measure (47) describes the amount of entanglement produced by the reduced
density operator (46) defined for $p$ particles from a quantum ensemble of
$N$ particles.

\subsection{Statistical mechanics}

Statistical properties of multiparticle systems are characterized by
reduced density matrices [41]. Therefore, it is natural to consider
entanglement realized by these matrices. The general scheme of measuring
entanglement production can be as follows.

Let $\cX=\{ x\}$ be a characteristic space of physical coordinates $x$,
whose concrete nature can be arbitrary. For example, $x$ can be a set of
spatial Cartesian coordinates, or it may be a set of momentum variables,
or a set of quantum numbers or mode indices. Assume that there exist several
such characteristic spaces $\cX_i=\{ x_i\}$, with $i=1,2,\ldots p$. And
let each space $\cX_i$ be measurable, with a differential measure $dx_i$
allowing for defining the Lesbegue integration over $\cX_i$. In the case of
discrete variables, the differential measure is atomic, so that integration
reduces to summation. The total characteristic space is the direct product
$$
\cX \equiv {\times}_{i=1}^p \; \cX_i \qquad (p=1,2,\ldots, N) \; .
$$
The set $x^p\equiv\{ x_1,x_2,\ldots,x_p\}$ is an element of $\cX=\{ x^p\}$.
The space $\cX$ is measurable, with a differential measure $dx^p\equiv
\prod_{i=1}^p dx_i$.

Let the elements of the single-partite Hilbert space $\cH_i$ be the vectors
$\vp_i=[\vp_i(x_i)]$ treated as columns with respect to the variable $x_i$,
with the scalar product
$$
\vp_i^+\vp_i' = (\vp_i,\vp_i') =
\int \vp_i^*(x_i)\; \vp_i'(x_i) \; dx_i \; .
$$
A $p$-order reduced density matrix is defined as a matrix
\be
\label{49}
\rho_p = \left [ \rho_p(x^p,y^p)\right ]
\ee
with respect to $x^p$ and $y^p$, whose elements are
\be
\label{50}
\rho_p(x^p,y^p) \equiv {\rm Tr}_\cF \psi(x_1)\ldots \psi(x_p) \hat\rho
\psi^\dgr(y_p) \ldots \psi^\dgr(y_1) \; ,
\ee
where $\psi(x)$ is a field operator, $\hat\rho$ is a statistical operator,
and the trace is over the Fock space $\cF$. Definition (50) can equivalently
be written as the statistical average
$$
\rho_p(x^p,y^p) = \; <\psi^\dgr(y_p)\ldots \psi^\dgr(y_1) \psi(x_1)
\ldots \psi(x_p) > \; .
$$
By this definition, the matrix $\rho_p$ is self-adjoint semipositive. An
example is a first-order density matrix
$$
\rho_1^i = [\rho_1(x_i,y_i)] \qquad
(x_i,y_i \in \cX_i) \; ,
$$
with
$$
\rho_1(x,y) =\; <\psi^\dgr(y)\psi(x) > \; .
$$
A common name for $\rho_1^i$ is the single-particle density matrix, which, as
was emphasized above, does not imply a concrete physical particle but rather
tells what is the order of the reduced matrix.

The trace operations for the density matrices are given by the expressions
$$
{\rm Tr}_{\cH_i}\rho_1^i \equiv \sum_{n_i} \; < n_i|\rho_1^i|n_i>\; =
\int \rho_1(x_i,x_i) \;dx_i = N \; ,
$$
and similarly
\be
\label{51}
{\rm Tr}_\cH \rho_p = \int \rho_p(x^p,x^p) \; dx^p =
\frac{N!}{(N-p)!} \; ,
\ee
which yields the relation
\be
\label{52}
\rho_1^i =\frac{(N-p)!}{(N-1)!} \; {\rm Tr}_{\{\cH_{j\neq i}\} } \;
\rho_p \; .
\ee
For the product operator (19), we now have
\be
\label{53}
\rho_p^\otimes = \frac{N!}{(N-p)!\; N!} \; \otimes_{i=1}^p \rho_1^i \; ,
\ee
whose norm (14) reads
$$
||\rho_p^\otimes||_\cD = \frac{N!}{(N-p)!\; N^p}\; \prod_{i=1}^p
||\rho_1^i||_{\cH_i} \; .
$$

To effectively calculate the norms, entering the entanglement-production
measure (20), we need to specify the single-partite spaces $\cH_i$. Recall
that the latter can, generally, be chosen in an arbitrary way. However, for
density matrices, there exists a natural choice related to the eigenvectors
of $\rho_1^i$. These eigenvectors are termed {\it natural orbitals} [41].
Therefore, under $|n_i>$, we shall imply the eigenvectors of $\rho_1^i$. And
a closed linear envelope of these natural orbitals gives the {\it natural
single-partite} space $\cH_i$. Then one can write
$$
\rho_1^i = \sum_{n_i} \; D_{n_i n_i}^1 \; |n_i><n_i| \; .
$$
Keeping in mind the spectral norm, one has
$$
||\rho_1^i||_{\cH_i} = \sup_{n_i} |D_{n_i n_i}^1| \; .
$$
A $p$-order density matrix can be presented as an expansion
\be
\label{54}
\rho_p = \sum_{\{m_i n_i\} } \; D_{\{ m_i n_i\} }^p \;
|m_1\ldots m_p><n_1 \ldots n_p| \; ,
\ee
so that
$$
||\rho_p||_\cD = \sup_{\{ n_i\} } \; |D_{\{ n_i n_i\} }^p | \; .
$$
In the case of a system of identical particles, when all $\cH_i$ are just
copies of the same $\cH_1$, then
$$
||\rho_1^i||_{\cH_i} = ||\rho_1||_{\cH_1} \; .
$$
Therefore, the entanglement-production measure (20) becomes
\be
\label{55}
\ep(\rho_p) = \log \;
\frac{(N-p)!\; N^p\; ||\rho_p||_\cD}{N!\; ||\rho_1||^p_{\cH_1} } \; .
\ee
Thus, the problem of quantifying the entanglement generated among any $p$
particles from a given ensemble of $N$ particles is reduced to calculating
the norms of reduced density matrices.

\subsection{Evolutional entanglement}

The reduced density matrices (49), in general, depend on time,
\be
\label{56}
\rho_p(t) = [\rho_p(x^p,y^p,t) ] \; ,
\ee
which enters definition (50) either through the field operators $\psi(x,t)$
or through the statistical operator $\hat\rho(t)$. Consequently, the
entanglement-production measure (55) is, generally, also a function of time.
Temporal dependence of entanglement is named evolutional entanglement [4].

As an example of entanglement generated in a nonequilibrium system, let us
consider the case when the reduced density matrices have the structure of
the {\it mixed multimode state}
\be
\label{57}
\rho_p(t) = \frac{N!}{(N-p)!} \;
\sum_n w_n(t)\; |n\ldots n><n\ldots n| \; ,
\ee
where $w_n(t)$ are the fractional mode populations with the properties
$$
0 \leq w_n(t) \leq 1 \; , \qquad \sum_n w_n(t) = 1 \; .
$$
Such multimode, or in the simplest case two-mode, states can be created in
trapped Bose-Einstein condensates (see reviews [54--56]). This can be done,
e.g., by separating a Bose condensate in a two-well, or a multiwell potential
[55,56]. Another possibility is by generating topological coherent modes in
a trapped condensate, by means of resonant alternating fields [48--50]. The
mixed atom-molecule condensate can also be treated as a two-mode coherent
system [57]. Coherent collisions between matter waves result in the formation
of an effective multimode system [58]. A similar system can also be created
in the process of superradiant scattering of an atomic Bose-condensed cloud
[59--61]. A multimode Bose-Einstein condensate has many analogies with
coherent optical systems and, in particular, with lasers [50,56,62].

For the density matrix (57), we get the norm
\be
\label{58}
||\rho_p(t)||_\cD = \frac{N!}{(N-p)!} \; \sup_n w_n(t) \; ,
\ee
while for the first-order matrix (52), we have
$$
||\rho_1^i(t)||_{\cH_i} = N\; \sup_n w_n(t) \; .
$$
Thus, for the product operator (53), we find
\be
\label{59}
||\rho_p^\otimes(t)||_\cD = \frac{N!}{(N-p)!} \; \sup_n\; w_n^p(t) \; .
\ee
As a result, the entanglement-production measure is
\be
\label{60}
\ep(\rho_p) =  (1-p)\log\; \sup_n \; w_n(t) \; .
\ee
The temporal behaviour of the fractional mode populations $w_n(t)$ are
defined by the evolution equations describing the corresponding process.
For instance, in the case of Bose-Einstein condensates, the evolution
equations for the mode populations follow from the time-dependent
Gross-Pitaevskii equation [48--50,62]. The fractional mode populations
$w_n(t)$, depending on the physical situation considered, vary in time
between $0$ and $1$. The maximal entanglement production happens at that
time when all populations coincide, so that $w_n(t)=1/m$, where
$m\equiv\sum_n 1$. That is, the entanglement-production measure (60) can
vary in the interval
$$
0\leq \ep(\rho_p) \leq (p-1)\log m \; .
$$

Thus, the entanglement-production measure (60) depends on time through the mode
populations. The evolution of the latter in some cases can be regulated. For
example, when the condensate mode structure is due to the resonant generation
of topological coherent modes, the temporal behaviour of mode populations is
governed by the applied alternating fields [48--50]. In this way, the time
evolution of the entanglement-production measure (60) can be controlled, which
opens wide possibility for information processing.

\subsection{Spin entanglement}

Entanglement production, being defined for arbitrary operators, can be
considered for systems of any nature. An important class of systems is that
corresponding to spin ensembles. Spin entanglement can be studied in the same
way as entanglement of particles, by invoking a kind of density matrices.

Let spin operators $\bS_i=\{ S_i^\al\}$, with $\al=x,y,z$, be associated
with a lattice $\Bbb{Z}_N=\{\ba_i\}$ whose lattice sites are enumerated by
an index $i=1,2,\ldots,N$. Similarly to the reduced density matrices (49),
composed of field operators, we may introduce [40] {\it spin density matrices}
\be
\label{61}
R_p = \left [ R_{\{ ij\} }^{\{\al\bt\} } \right ] \; ,
\ee
with the elements
\be
\label{62}
R_{\{ ij\} }^{\{\al\bt\} } \equiv \; < S_{j_p}^{\bt_p} \ldots
S_{j_1}^{\bt_1} S_{i_1}^{\al_1} \ldots S_{i_p}^{\al_p} >
\ee
composed of statistical averages of spin operators. Form (61) is treated
as a matrix with respect to all indices $\{ ij\}$ as well as $\{ \al\bt\}$,
although in principle, one could also consider the matrices $R_p^{\{\al\bt\}}$
and $R_{p\{ ij\}}$ defined as matrices with respect to only the indices
$\{ ij\}$ or $\{\al\bt\}$, under fixed other indices. The first-order spin
density matrix (61) is $R_1=[R_{ij}^{\al\bt}]$, whose elements are the
correlation functions
\be
\label{63}
R_{ij}^{\al\bt} \equiv \; < S_j^\bt\; S_i^\al> \; .
\ee
Matrices (61) are self-adjoint, since such are the spin operators. With the
traces
$$
{\rm Tr}_{\cH_1}\; R_1 \equiv \sum_i \sum_\al \; < S_1^\al\; S_i^\al> \;
= NS (S+1) \; , \qquad
{\rm Tr}_\cH\; R_p = (NS)^p\; (S+1)^p \; ,
$$
where $S$ is the maximal quantum number of each spin, the product operator
(19) becomes
\be
\label{64}
R_p^\otimes = R_1\otimes R_1 \otimes \ldots \otimes R_1 \; .
\ee

To calculate the operator norms, the single-partite space $\cH_1$ can be
specified as the closed linear envelope of {\it natural spin orbitals},
which are the eigenfunctions of $R_1$. This is in analogy with the case of
reduced density matrices $\rho_p$. Finally, the spin entanglement measure
(20) takes the form
\be
\label{65}
\ep(R_p) = \log\; \frac{||R_p||_\cD}{||R_p^\otimes||_\cD} \; .
\ee
Explicit calculation of this measure will be illustrated in the following
sections.

\section{Statistical Thermal Entanglement}

Instead of considering entanglement realized by reduced density matrices,
as in the previous section, one may study entanglement produced by a
statistical operator $\hat\rho$. The latter for a statistical system in
thermal equilibrium reads
\be
\label{66}
\hat\rho = \frac{1}{Z} \; e^{-\bt H} \; , \qquad
Z \equiv {\rm Tr} \; e^{-\bt H} \; ,
\ee
where $\bt\equiv(k_BT)^{-1}$, with $k_B$ being the Boltzmann constant and
$T$, temperature. One can define, in complete analogy with the previous
sections, the reduced operators by tracing out some of the states of the
total space $\cH$. For example,
\be
\label{67}
\hat\rho_1^i \equiv {\rm Tr}_{\{\cH_{j\neq i}\} } \; \hat\rho \; .
\ee
Entanglement, related to statistical thermal operators, is usually
characterized by the concurrence between a pair of qubits [63--68]. Here
we show that the entanglement-production measure (20) provides a natural
characteristic of the thermal entanglement, which is produced by the
statistical operator (66).

Let us consider the Ising model
\be
\label{68}
H = -\; \frac{1}{2}\; \sum_{i\neq j}^N \; J_{ij}\; S_i^z \; S_j^z -
B\; \sum_{i=1}^N\; S_i^z \; ,
\ee
with an exchange interaction $J_{ij}=J_{ji}$ and an external magnetic field
$B\geq 0$. Positive interaction $J_{ij}>0$ is called ferromagnetic, while
negative $J_{ij}<0$, antiferromagnetic. It is natural to define the basis
states $|n_i>$ as the eigenvectors of $S_i^z$. Then the single-partite space
$\cH_i$ is the span of the basis $\{|n_i>\}$. For concreteness, let us take
a two-site case, denoting $J\equiv J_{12}$, and let us consider spin-one-half
operators $S_i^z$, with $S=1/2$. Introduce the dimensionless coupling $g$ and
magnetic field $b$ by the expressions
\be
\label{69}
g \equiv \bt\; J\; S^2\; , \qquad b \equiv \bt\; B \; .
\ee
Then we have
\be
\label{70}
\hat\rho_2 = \frac{1}{Z}\; \exp\left\{ 4 g S_1^z\; S_2^z + b\left (
S_1^z + S_2^z \right ) \right \} \; ,
\ee
where
$$
Z=2\left ( e^g{\rm cosh} b + e^{-g}\right ) \; .
$$
For the reduced operator
\be
\label{71}
\hat\rho_1^i = {\rm Tr}_{\cH_{j\neq i}} \; \hat\rho_2 \; ,
\ee
we find
\be
\label{72}
\hat\rho_1^i = \frac{1}{Z}\; \exp\left\{ 2gS_i^z + b \left ( S_i^z +
\frac{1}{2}\right ) \right \} + \frac{1}{Z}\; \exp\left\{ -2g S_i^z +
b\left ( S_i^z -\; \frac{1}{2}\right ) \right \} \; .
\ee
The disentangled set $\cD$ consists of the states $|\uparrow\uparrow>$,
$|\uparrow\downarrow>$, $|\downarrow\uparrow>$ and $|\downarrow\downarrow>$.
For the norms of $\hat\rho_2$ and $\hat\rho_1^i$, we get
\be
\label{73}
||\hat\rho_2||_\cD = \frac{1}{Z}\; \sup\left\{ e^{g+b},\; e^{-g}\right\}
\ee
and, respectively,
\be
\label{74}
||\hat\rho_1^i||_\cD = \frac{1}{Z}\; \left ( e^{g+b} + e^{-g}\right ) \; .
\ee
This, taking into account that $\hat\rho_2^\otimes=\hat\rho_1^1\otimes
\hat\rho_1^2$, results in the entanglement-production measure
\be
\label{75}
\ep(\hat\rho_2) = \log\left [
\frac{2(1+e^{2g}{\rm cosh}b)}{(1+e^{b+2g})^2}\; \sup\left\{ 1, \;
e^{b+2g}\right\} \right ] \; ,
\ee
describing the pairwise spin entanglement production.

Analyzing the properties of measure (75), it is interesting to compare its
behaviour with that of the average magnetization per spin
$$
M \equiv -\; \frac{1}{N}\; \frac{\partial F}{\partial B} \; , \qquad
F \equiv -\; \frac{1}{\bt}\; \ln Z \; ,
$$
for which one has
\be
\label{76}
M = \frac{e^{2g}{\rm sinh} b}{2(1 + e^{2g}{\rm cosh}b)} \; .
\ee
Expressions (75) and (76) are functions of two variables, coupling
$g\in(-\infty,+\infty)$ and magnetic field $b\in[0,\infty)$.

In the case of zero magnetic field, when $B\ra 0$ and $b\ra 0$, one
has $M\ra 0$ and
\be
\label{77}
\lim_{b\ra 0}\; \ep(\hat\rho_2) =
\log \; \frac{e^{|g|}}{{\rm cosh}g} \; .
\ee
For low temperature, when $T\ra 0$, hence $\bt\ra\infty$ and $g\ra\pm\infty$,
we get
\be
\label{78}
\lim_{g\ra\pm\infty} \; \lim_{b\ra 0} \; \ep(\hat\rho_2) = \log 2 \; ,
\ee
that is, the maximal pairwise entanglement production. In the opposite case of
high temperatures, when $T\ra\infty$, so that $\bt\ra 0$ and $g\ra 0$, we find
\be
\label{79}
\lim_{g\ra 0}\; \lim_{b\ra 0} \; \ep(\hat\rho_2) = 0 \; ,
\ee
which means that high-temperature fluctuations destroy entanglement production.
Note that the limits $b\ra 0$ and $g\ra 0$ commute with each other.

If the external magnetic field increases, with $B\ra\infty$ and $b\ra\infty$,
then
\be
\label{80}
\lim_{b\ra\infty}\; \ep(\hat\rho_2) = 0 \; , \qquad
\lim_{b\ra\infty} M = \frac{1}{2} \; .
\ee
This shows that there is no entanglement production between perfectly aligned
spins.

When temperature diminishes, such that $T\ra 0$ and $\bt\ra\infty$, but
the magnetic field is nonzero, $B\neq 0$, then the resulting expressions
depend on the relation between $b\ra\infty$ and $|g|\ra\infty$. Thus, we
obtain
\begin{eqnarray}
\lim_{T\ra 0} \; \ep(\hat\rho_2) =\left \{
\begin{array}{ll}
\label{81}
log 2 & (b+2g\ra -\infty ) \\
log(3/4) & (b+2g\ra 0) \\
0 & (b+2g\ra +\infty ) .
\end{array} \right.
\end{eqnarray}
And for magnetization (76), we find
\begin{eqnarray}
\lim_{T\ra 0} \; M =\left \{
\begin{array}{ll}
\label{82}
0 & (b+2g\ra -\infty ) \\
1/6 & (b+2g\ra 0) \\
1/2 & (b+2g\ra +\infty ) .
\end{array} \right.
\end{eqnarray}

Magnetization plays the role of an order parameter. The above analysis
shows that the larger is the magnetization, the smaller is the
entanglement-production measure. In this case, entanglement production and
order are complimentary to each other. To illustrate this by another example,
let us consider the Hamiltonian (68) with ferromagnetic interactions
$J_{ij}>0$ of long-range type, when $J_{ij}$ depends on $N$ so that
\be
\label{83}
\lim_{N\ra\infty} J_{ij} = 0 \; , \qquad \lim_{N\ra\infty}
\frac{1}{N}\; \sum_{i\neq j}^N J_{ij} < \infty \; .
\ee
In that case, as is known [69], the Hamiltonian (68) is asymptotically, as
$N\ra\infty$, equivalent to the mean-field form $\sum_i H_i$, where
$$
H_i = -\left ( \sum_j J_{ij}\; < S_j^z>\; + \; B\right ) S_i^z +
\frac{1}{2}\; \sum_j J_{ij}\; < S_i^z><S_j^z> \; .
$$
This implies that the statistical operator (66) is asymptotically, as
$N\ra\infty$, equivalent to
\be
\label{84}
\hat\rho_N \ra \otimes_{i=1}^N \; \hat\rho_1^i \; , \qquad
\hat\rho_1^i = \frac{1}{Z^{1/N}}\; e^{-\bt H_i} \; .
\ee
But then $\hat\rho_N\ra\hat\rho_N^\otimes$ and we come to the limit
\be
\label{85}
\lim_{N\ra\infty} \ep(\hat\rho_N) = 0 \; .
\ee
Long-range ferromagnetic interactions, satisfying condition (83), organize
long-range magnetic order in the spin system with a finite magnetization
$M$. At the same time, this yields the absence of entanglement production
between ferromagnetically aligned spins.

\section{Entanglement and Phase Transitions}

The analysis of the previous section hints that there should be a relation
between the entanglement-production measure and an order parameter. The
latter, in turn, experiences dramatic changes at phase transitions. Hence,
entanglement production may also exhibit essential changes under phase
transformations, which is illustrated below.

\subsection{Bose-Einstein condensation}

Let us study the entanglement realized by the reduced density matrices (49).
The properties of the latter are described in detail in book [41]. At high
temperature, much larger than the condensation temperature $T_c$, one has
\be
\label{86}
||\rho_p||_\cD \simeq ||\rho_1||^p_{\cH_1} \qquad (T\gg T_c) \; .
\ee
Therefore, the entanglement-production measure (55) is
\be
\label{87}
\ep(\rho_p) \simeq \log\; \frac{(N-p)!\; N^p}{N!} \qquad (T\gg T_c) \; ,
\ee
which is typical of the Hartree-Fock form.

At low temperature $T\ll T_c$, we find [41] that
\be
\label{88}
||\rho_p||_\cD \simeq \; \frac{N!}{(N-p)!} \; , \qquad
||\rho_1||_{\cH_1} \simeq N \; .
\ee
Therefore, measure (55) becomes
\be
\label{89}
\ep(\rho_p) \simeq 0 \qquad (T\ll T_c) \; .
\ee
This means that entanglement production diminishes when the Bose-Einstein
condensation occurs.

\subsection{Superconducting transition}

At temperatures much higher that the critical temperature $T_c$, density
matrices are of the Hartree-Fock type, which, for large $N\gg 1$, yields
\be
\label{90}
\ep(\rho_p) \simeq \frac{p(p-1)}{2N}\; \log e \qquad (T\gg T_c) \; .
\ee
At temperatures below $T_c$, the structure of the reduced density matrices
essentially changes, as is thoroughly  described in Ref. [41]. Then one has
\begin{eqnarray}
\label{91}
||\rho_p||_\cD \simeq c_p \times\left\{
\begin{array}{ll}
N^{(p-1)/2} & \;\; \; (p\; odd) \\
N^{p/2} & \; \; \; (p\; even) \; ,
\end{array}\right.
\end{eqnarray}
where $c_p$ is a constant of order one. From here,
$$
||\rho_1||_{\cH_1} \simeq c_1 \; , \qquad ||\rho_p^\otimes||_\cD \simeq
\frac{N!\; c_1^p}{(N-p)!\; N^p} \; .
$$
Thus, for $T<T_c$, finite $p$, and large $N\gg 1$, we obtain
\begin{eqnarray}
\label{92}
\ep(\rho_p) \simeq \left\{
\begin{array}{ll}
\frac{p-1}{2}\; \log N & \;\; \; (p\; odd) \\
\frac{p}{2}\; \log N & \; \; \; (p\; even) \; .
\end{array}\right.
\end{eqnarray}
Comparing Eqs. (90) and (92), we see that the entanglement-production
measure $\ep(\rho_p)$ increases under arising superconductivity. This is
contrary to what happens under Bose-Einstein condensation. Such a difference
should not be surprising and it can be easily understood as follows.
Considering here entanglement production, the single-partite space $\cH_1$
has been treated as the space of single-particle quantum states. In the case
of the Bose-Einstein condensation, there appears ordering of particles,
which leads to the decrease of their entanglement production. But under
superconducting transition, the arising order has to do with pairs of particles,
that is, Cooper pairs, and not with separate particles. If we define the
single-partite space as the space of the Cooper pairs quantum states, then
the overall situation would become similar to that occurring at Bose-Einstein
condensation. Then the order appearing between Cooper pairs under
superconducting transition would result in the diminishing entanglement
production of these pairs.

\subsection{Magnetic transition}

To study the interplay between magnetic ordering and entanglement production,
we shall invoke the spin density matrices (61). As a particular case, let us
consider these spin density matrices composed of the $z$-components $S_i^z$
of spin operators, with $i=1,2,\ldots,N$ enumerating lattice sites. Thus,
the {\it spin density matrix} $R_p=[R_{\{ ij\} }]$ possesses the elements
\be
\label{93}
R_{\{ ij\} } \equiv \;
< S_{j_p}^z \ldots S_{j_1}^z S_{i_1}^z \ldots S_{i_p}^z > \; .
\ee
Clearly, this is a Hermitian semipositive matrix.

The eigenfunctions of $R_1=[<S_j^z\; S_i^z>]$ are the vectors
\be
\label{94}
|k>\; = [\vp_k(\ba_i)] \; , \qquad \vp_k(\ba) \equiv \frac{1}{\sqrt{N}} \;
e^{i{\bf k}\cdot\ba} \; ,
\ee
treated as $N$-order columns with respect to the lattice vectors
$\ba_i\in\Bbb{Z}_N$. Functions (94) form a complete orthonormal basis
$\{|k>\}$. A single-partite space is defined as the closed linear envelope
$\cH_1\equiv\overline\cL\{|k>\}$. The norm of $R_p$ over $\cD$ can be
calculated as
\be
\label{95}
||R_p||_\cD =\sup_{\{ k_i\} } \; < k_1 \ldots k_p|R_p|k_1\ldots k_p> \; .
\ee
The elements of $R_1$ are the correlation function $<S_i^z\; S_j^z>$ having the
property [70]
\be
\label{96}
\lim_{a_{ij}\ra\infty} \; <S_i^z\; S_j^z> \; \equiv M^2 \; ,
\ee
where $a_{ij}\equiv|\ba_i-\ba_j|$ and $M$ is the magnetization
$$
M\equiv \frac{1}{N} \; \sum_{i=1}^N \; < S_i^z> \; .
$$
Here we keep in mind ferromagnetic phase transition. Calculating norms
(95), we use, for simplicity, the mean-field approximation and assume large
$N\ra\infty$. Above the critical temperature $T_c$, where $M=0$, we find
\be
\label{97}
||R_p||_\cD = (2p-1)!!\; S^{2p} \qquad (T\geq T_c) \; ,
\ee
with
$$
(2p-1)!! \equiv 1\times 3 \times 5\times \ldots\times (2p-1) =
\frac{(2p)!}{2^p\;p!} \; .
$$
And below $T_c$, where $M\neq 0$, we have
\be
\label{98}
||R_p||_\cD \simeq N^p\; M^{2p} \qquad (T< T_c) \; .
\ee
In particular,
$$
|| R_1||_{\cH_1} = S^2 + NM^2
$$
at all temperatures. Therefore, for the entanglement-production measure (65),
we obtain
\be
\label{99}
\ep(R_p) = \log\; \frac{(2p)!}{2^p\; p!} \qquad (T\geq T_c)
\ee
above the transition temperature. For instance,
$$
\ep(R_2) = \log 3 \; , \qquad \ep(R_3) = \log 15 \; .
$$
And below $T_c$, where there appears magnetic order, we get
\be
\label{100}
\ep(R_p) \simeq 0 \qquad (T< T_c)
\ee
for any finite $p$.

As is seen, the arising magnetic order also leads to the decrease of
entanglement production, similarly to what happens at Bose-Einstein
condensation. The same qualitative relation between entanglement production
and ordering for different phase transitions can be understood if to keep
in mind that a ferromagnetic phase transition is accompanied by the
condensation of magnons [70].

\subsection{Order indices}

The intimate relation between entanglement production and ordering,
illustrated above by several examples, can be explained by the most general
arguments. For this purpose, we need to resort to the notion of operator
order indices that can be introduced for arbitrary operators [40], and, in
particular, for reduced density matrices [41]. The {\it operator order index}
of an operator $A$ is defined [40] as
$$
\om(A) \equiv \frac{\log ||A||}{\log|{\rm Tr}\; A|} \; .
$$
In describing the ordering in physical systems, the role of operators $A$
is played by reduced density matrices. The latter can be the standard density
matrices constructed of field operators [41], as well as spin density matrices,
lattice density matrices, or matrices composed of other operators [40]. The
larger is the order index $\om(A)$,  the higher is the level of ordering
corresponding to the operator $A$. Such order indices characterize both
long-range as well as various types of mid-range order. They are suitable
for describing off-diagonal and diagonal orders and can be applied for any
physical system, equilibrium or nonequilibrium, infinite or finite. Also, they
do not involve the notion of broken symmetry and can be employed when the order
parameters are not defined.

If the operator $A$ represents a density matrix, it is semipositive.
Then $||A||\leq{\rm Tr}A$, because of which $\om(A)\leq 1$. Assume that a
semipositive operator $A$ is associated with a kind of ordering in a physical
system. The order is absent when $||A||\ll{\rm Tr}A$, and then $\om(A)\ll 1$.

There are two types of long-range order that may develop in physical systems,
total and even orders [40,41]. Under the arising {\it total long-range order},
the norms of the related density matrices increase, so that $||\rho_p||\sim
{\rm Tr}\rho_p$, hence $\om_p(\rho)\ra 1$. The order is total, which implies
that the property $||\rho_p||\sim{\rm Tr}\rho_p$ is valid for the density
matrices of all orders $p=1,2,\ldots$, including $||\rho_1||\sim{\rm Tr}
\rho_1$. From here, $||\rho_p^\otimes||\sim{\rm Tr}\rho_p^\otimes$. According
to the normalization condition (18), we have ${\rm Tr}\rho_p={\rm Tr}
\rho_p^\otimes$. Therefore, $||\rho_p||\sim ||\rho_p^\otimes||$, which
results in $\ep(\rho_p)\ra 0$. As is shown above, this situation takes place
at Bose-Einstein condensation and ferromagnetic transition. Thus, under the
appearing total long-range order, the order indices increase but the
entanglement-production measure decreases,
$$
\om(\rho_p)\ra 1 \; , \qquad \ep(\rho_p)\ra 0 \qquad (total\; order) \; .
$$
Here increasing total order implies diminishing entanglement production.

The situation is different for {\it even long-range order}. Then the norms of
the density matrices also increase, but so that [40,41]
\begin{eqnarray}
||\rho_p|| \sim \left\{
\begin{array}{ll}
\sqrt{{\rm Tr}\rho_p/N} & \;\;\; (p\; odd) \\
\nonumber
\sqrt{{\rm Tr}\rho_p} & \;\;\; (p\; even) \; .
\end{array}\right.
\end{eqnarray}
For instance, $||\rho_1||\sim const$ but $||\rho_2||\sim N$. Because of
this, the order indices are different for $p$ odd and $p$ even,
\begin{eqnarray}
\om(\rho_p) \ra \left\{
\begin{array}{ll}
(p-1)/2p & \;\;\; (p\; odd) \\
\nonumber
1/2 & \;\;\; (p\; even) \; .
\end{array}\right.
\end{eqnarray}
There is no order in the single-particle matrix $\rho_1$, as far as
$\om(\rho_1)=0$, but $\om(\rho_2)=1/2$. For all $p=1,2,\ldots$, we have
$\om_{2p-1}<\om_{2p}$, and they coincide only in the limit $p\ra\infty$. Such
a behaviour is typical of superconducting transition. Entanglement production
is also different for odd and even numbers of particles,
\begin{eqnarray}
\ep(\rho_p) \ra \left\{
\begin{array}{ll}
\frac{p-1}{2}\;\log N & \;\;\; (p\; odd) \\
\nonumber
\frac{p}{2}\;\log N & \;\;\; (p\; even) \; .
\end{array}\right.
\end{eqnarray}
The entanglement-production measure increases, since the inequality
$||\rho_1||\ll ||\rho_p||$, with $p>1$, yields
$||\rho_p^\otimes||\ll||\rho_p||$. Now both the order indices for
$p>1$ and entanglement-production measure increase together. Hence,
increasing even order results as well in increasing entanglement
production.

\vskip 3mm

In conclusion, a general definition of entanglement-production measure
for arbitrary operators is introduced. The concept is valid for systems
of any nature. The interplay between entanglement production and phase
transitions is elucidated.

\end{document}